\documentclass[reprint, twocolumn, superscriptaddress,prb]{revtex4-2}
\usepackage{bm, amsmath, amsfonts, amssymb, braket}
\usepackage{subfigure}
\usepackage{color}
\usepackage{graphicx}
\usepackage{ulem}
\usepackage{dcolumn} 

\usepackage{rotating} 

\begin{document}
\title{Relationship between two-particle topology and fractional Chern insulator}
\author{Nobuyuki Okuma}
\email{okuma@hosi.phys.s.u-tokyo.ac.jp}
\affiliation{%
 Center for Gravitational Physics and Quantum Information, Yukawa Institute for Theoretical Physics, Kyoto University, Kyoto 606-8502, Japan
}%

\author{Tomonari Mizoguchi}
\affiliation{Department of Physics, University of Tsukuba, Tsukuba, Ibaraki 305-8571, Japan}
\email{mizoguchi@rhodia.ph.tsukuba.ac.jp}

\date{\today}
\begin{abstract}
Lattice generalizations of fractional quantum Hall (FQH) systems, called fractional Chern insulators (FCIs), have been extensively investigated in strongly correlated systems.
Despite many efforts, previous studies have not revealed all of the guiding principles for the FCI search.
In this paper, we investigate a relationship between the topological band structure in the two-particle problem and the FCI ground states in the many-body problem.
We first formulate the two-particle problem of a bosonic on-site interaction projected onto the lowest band of a given tight-binding Hamiltonian.
We introduce a reduced Hamiltonian whose eigenvalues correspond to the two-particle bound-state energies.
By using the reduced Hamiltonian, we define the two-particle Chern number and numerically check the bulk-boundary correspondence that is predicted by the two-particle Chern number.
We then propose that a nontrivial two-particle Chern number of dominant bands roughly indicates the presence of bosonic FCI ground states at filling factor $\nu=1/2$.
We numerically investigate this relationship in several tight-binding models with Chern bands 
and find that it holds well in most of the cases, albeit two-band models being exceptions.
Although the two-particle topology is neither a necessary nor a sufficient condition for the FCI state as other indicators in previous studies, our numerical results indicate that the two-particle topology characterizes the degree of similarity to the FQH systems.

\end{abstract}
\maketitle
\section{Introduction}
The physics of topological phases has received much attention in recent years.
Among them, much is known theoretically about the topological insulators~\cite{Kane-review,Zhang-review} because their essence can be understood in terms of a one-particle picture.
The integer Quantum Hall (IQH) system \cite{Klitzing-80} is one of the first examples of topological insulators.  
Although the bulk of this system is a band insulator in which an integer number of Landau levels are fully occupied, 
the gapless edge states that are protected by bulk topology lead to a quantized Hall conductance.
This topological protection, called the bulk-boundary correspondence~\cite{Harperin-82,Hatsugai-93-PRB,Hatsugai-93-PRL}, is a central property of the topological insulators.

When the Landau levels are partially filled with the fractional filling factor $\nu=p/q$,
the interplay between bulk topology and strong correlation leads to the fractional quantum Hall (FQH) effect~\cite{FQHE-exp-82,Laughlin-Wavefunction-83,yoshioka-textbook-02,halperin-fractional-textbook-20}.
Mathematically, the FQH system is classified as an intrinsic topological order \cite{wen1990topological}, 
which can not be understood in terms of a one-particle picture.
While the topological insulators do not host bulk degrees of freedom, the systems in topological order host fractional excitations in the bulk \cite{Laughlin-anyon-83}, called anyons \cite{leinaas1977theory,wilczek1982quantum}.
Depending on the filling factor, there exist the Abelian \cite{Laughlin-anyon-83} and non-Abelian anyons \cite{moore1991nonabelions, Nayak-08-review} in the FQH systems.
The topological orders are classified by the types of anyons, 
and the emergence of anyons is accompanied by many exotic properties such as topological long-range entanglement \cite{Kitaev-Preskill-06, Levin-Wen-06} and ground-state degeneracy \cite{Wen-Niu-90}. 
These concepts have applications in topological quantum computation, which is expected to realize fault tolerance against quantum errors~\cite{Nayak-08-review}.

In the physics of IQH systems, 
a lattice construction without a magnetic field was firstly given by Haldane~\cite{Haldane-88}.
In the Haldane model, a nontrivial Chern number of a dispersive band mimics a Landau level and topologically protects the gapless edge states. Nowadays, a lattice system with a nontrivial Chern band is called the Chern insulator.
Similarly, lattice implementations of FQH ground states, called fractional Chern insulators (FCIs) \cite{Regnault-Bernevig-11, Bergholtz-Liu-13, Parameswaran-13, Liu-Bergholtz-review-22}, have been extensively studied for interacting Hamiltonians with a fractional filling whose non-interacting part is a Chern insulator.
However, it is not easy to determine whether or not the FCI states become the ground states of a given model.
In the physics of FQH systems, the flatness of the single-particle energy and analytical simplicity of the wave function of the Landau levels greatly simplify the problem, and the ground state is well described by a simple analytical many-body wave function, 
called the Laughlin wave function \cite{Laughlin-Wavefunction-83}.
Therefore, one guiding principle for the search of the FCI states is to focus on the one-particle properties and find a Chern band that mimics the Landau levels.
In this context, the flatness of the Chern band is the simplest criterion because it enhances the correlation effect within the Chern band.
A more nontrivial task is to find the conditions on the Bloch wave functions. By focusing on the properties in FQH systems such as the Girvin-MacDonald-Platzman algebra \cite{Girvin-MacDonald-Platzman-86}, 
previous studies investigated the importance of momentum-space distributions of the Berry curvature and quantum metric \cite{Parameswaran-Roy-Sondhi-12,Parameswaran-13,Roy-geometry-14,Jackson-Moller-Roy-15,Claassen-Lee-Thomale-Qi-Devereaux-15,Lee-Claassen-Thomale-17,Mera-Ozawa-21-2,Varjas-Abouelkomsan-Kang-Bergholtz-22}.
Indicators for the FCI ground states other than  the one-particle properties have also been studied.
Like the Cooper problem in the theory of superconductivity \cite{Cooper-56}, the two-particle problem gives significant insight in the FCI problem.
References \cite{Lauchli-Liu-Bergholtz-Moessner-13,Liu-Bergholtz-Kapit-13} pointed out a rough correspondence between Haldane's pseudopotential \cite{Haldane-Hierarchy-83} in FQH problem and the two-particle band structure.

Despite many efforts, previous studies have not revealed all of the guiding principles for the FCI search.
In this paper, we investigate a relationship between the two-particle topology and the FCI phase. 
We consider a bosonic on-site interaction Hamiltonian projected onto the lowest band of a given tight-binding model.
In the first half of this paper, we formulate the two-particle problem, in which the finite-energy eigenspace consists of the bound states, and define the reduced Hamiltonian that extracts only the bound-state information.
By using the reduced Hamiltonian, we define the two-particle Chern number for two-particle bands and numerically check the bulk-boundary correspondence that is predicted by the two-particle Chern number.
In the second half, we propose that a nontrivial two-particle Chern number ($|C|=1$) of the dominant two-particle bands roughly indicates the presence of the FCI ground states at the filling factor $\nu=1/2$. We numerically check this rough correspondence between the two-particle topology and the presence or absence of FCI ground states. Although the two-particle topology is neither a necessary nor a sufficient condition for the FCI state as other indicators in previous studies, the numerical results indicate that the two-particle topology characterizes the degree of similarity to the FQH system.

This paper is organized as follows.
In Sec. \ref{model-and-convention}, we introduce several conventions and define the projected bosonic Hamiltonian that is of interest.
In Sec. \ref{formalism}, we give the matrix representation of the projected Hamiltonian with only two particles and define the reduced Hamiltonian that is convenient for the topological analysis.
In Sec. \ref{two-particle-topology}, we propose a relationship between the two-particle topology in the two-particle problem and the FCI phase in the many-body problem at $\nu=1/2$.
In Sec. \ref{sec:various_model}, we numerically check the relationship that is proposed in Sec. \ref{two-particle-topology}.
In Sec. \ref{summary-and-discussions}, we summarize the paper and mention several future works.

\section{Model and convention\label{model-and-convention}}
We consider a bosonic, two-dimensional lattice Hamiltonian with repulsive on-site Hubbard interaction:
\begin{align}
    &H_{\rm full}=H_0+H_{\rm int}\notag\\
    =&\sum_{\bm{R},\bm{R}',i,i'}t_{\bm{R}i,\bm{R}'i'}c^{\dagger}_{\bm{R},i}c_{\bm{R}',i'}+U\sum_{\bm{R},i}n_{\bm{R},i}(n_{\bm{R},i}-1)\notag\\
    =&\sum_{\bm{R},\bm{R}',i,i'}t_{\bm{R}i,\bm{R}'i'}c^{\dagger}_{\bm{R},i}c_{\bm{R}',i'}+U\sum_{\bm{R},i}c^{\dagger}_{\bm{R},i}c^{\dagger}_{\bm{R},i}c_{\bm{R},i}c_{\bm{R},i},
\end{align}
where a matrix $t$ represents the hopping term with discrete translation invariance, $U>0$ is the strength of the interaction, $n$ is the local bosonic number operator, and $(c,c^{\dagger})$ are bosonic creation and annihilation operators.
In the following, we set $U=1$.
$\bm{R}$ and $i$ label the Bravais lattice and the orbital and/or sublattice degrees of freedom in a unit cell, respectively. 

When we impose the periodic boundary condition (PBC) both in the $x$ and $y$ directions, it is convenient to introduce the momentum-space picture.
By performing the Fourier transform, one can rewrite the quadratic Hamiltonian as
\begin{align}
    H_0=\sum_{\bm{k},i,j} H_{i,j}(\bm{k})c^{\dagger}_{\bm{k},i}c_{\bm{k},j},
\end{align}
where the matrix $H(\bm{k})$ is the Bloch Hamiltonian at crystal momentum: 
\begin{align}
    \bm{k}  
    =& (k_1,k_2) \notag \\
    =& \frac{n_1}{L_1}\bm{G}_1+\frac{n_2}{L_2}\bm{G}_2,~0\leq n_i~(\in\mathbb{Z}) \leq L_i-1 ~(i=1,2). \label{eq:k_discrite}
\end{align}
$\bm{G}_i$ and $L_i~(i=1,2)$ are the reciprocal lattice vector and the system size, respectively. 
When the unit cell has more than one orbital/sublattice degrees of freedom, there are two typical conventions for Fourier transform, depending on whether or not the relative positions $\bm{r}_i$ inside the unit cell are included in the Fourier factor.
We here adopt the Fourier factor without $\bm{r}_i$:
\begin{align}
    c^{\dagger}_{\bm{k},i}=\frac{1}{\sqrt{N_{\rm unit}}}\sum_{\bm{R}}e^{i\bm{k}\cdot\bm{R}}c^{\dagger}_{\bm{R},i},
\end{align}
where $N_{\rm unit}:=L_1L_2$ is the number of unit cells.
Under this convention, the Bloch Hamiltonian has the periodicity in momentum space:
\begin{align}
   H(\bm{k})=H(\bm{k} + \bm{G}_i).
\end{align}
By diagonalizing the Bloch Hamiltonian, one can rewrite the quadratic Hamiltonian $H_0$ as 
\begin{align}
    H_0=\sum_{\bm{k},a}\epsilon_{\bm{k},a}c^{\dagger}_{\bm{k},a}c_{\bm{k},a},
\end{align}
where $\epsilon_{\bm{k},a}$ is an eigenvalue of $H(\bm{k})$, and
\begin{align}
    &c^{\dagger}_{\bm{k},a}=\frac{1}{\sqrt{N_{\rm unit}}}\sum_{\bm{R},i}u_{\bm{k},a}(i)e^{i\bm{k}\cdot\bm{R}}c^{\dagger}_{\bm{R},i},\\
    &c_{\bm{R},i}=\frac{1}{\sqrt{N_{\rm unit}}}\sum_{\bm{k},a}u_{\bm{k},a}(i)e^{i\bm{k}\cdot\bm{R}}c_{\bm{k},a}\label{expansion}
\end{align}
with $\bm{u}_{\bm{k},a}$ being the corresponding unit eigenvector and $u_{\bm{k},a}(i)$ being the $i$th component of $\bm{u}_{\bm{k},a}$.
By using the expansion (\ref{expansion}), we can rewrite the interaction term as
\begin{align}
   & H_{\rm int}=\frac{U}{N_{\rm unit}}\sum_{\bm{q},\bm{k},\bm{k}'}\sum_{i}\sum_{a,b,c,d}\notag\\
    &u^{*}_{\bm{k},a}(i)u^{*}_{\bm{q}-\bm{k},b}(i)u_{\bm{q}-\bm{k}',c}(i)u_{\bm{k}',d}(i)
    c^{\dagger}_{\bm{k},a}c^{\dagger}_{\bm{q}-\bm{k},b}c_{\bm{q}-\bm{k}',c}c_{\bm{k}',d}.
\end{align}

In many theoretical studies of the FCIs, instead of the full Hamiltonian $H_{\rm full}$, the interaction Hamiltonian projected onto the lowest band $\alpha$ is of interest \cite{Bergholtz-Liu-13}:
\begin{align}
    H_{\rm proj}=\frac{U}{N_{\rm unit}}\sum_{\bm{q},\bm{k},\bm{k}'}\sum_{i}&u^{*}_{\bm{k},\alpha}(i)u^{*}_{\bm{q}-\bm{k},\alpha}(i)u_{\bm{q}-\bm{k}',\alpha}(i)u_{\bm{k}',\alpha}(i)\notag\\
    &c^{\dagger}_{\bm{k},\alpha}c^{\dagger}_{\bm{q}-\bm{k},\alpha}c_{\bm{q}-\bm{k}',\alpha}c_{\bm{k}',\alpha}.\label{projectedham}
\end{align}
This approximation is valid when the interaction is much smaller than the band gap and larger than the bandwidth of the band $\alpha$. 
These conditions are satisfied in the FQH system on the lowest Landau level \cite{yoshioka-textbook-02}.
In this paper, we consider the projected Hamiltonian (\ref{projectedham}) with $\alpha$ being a band with unit Chern number, $C=\pm1$. 

\section{Formalism: Two-particle Hamiltonian under projection\label{formalism}}
\begin{figure}[]
\begin{center}
 \includegraphics[width=8cm,angle=0,clip]{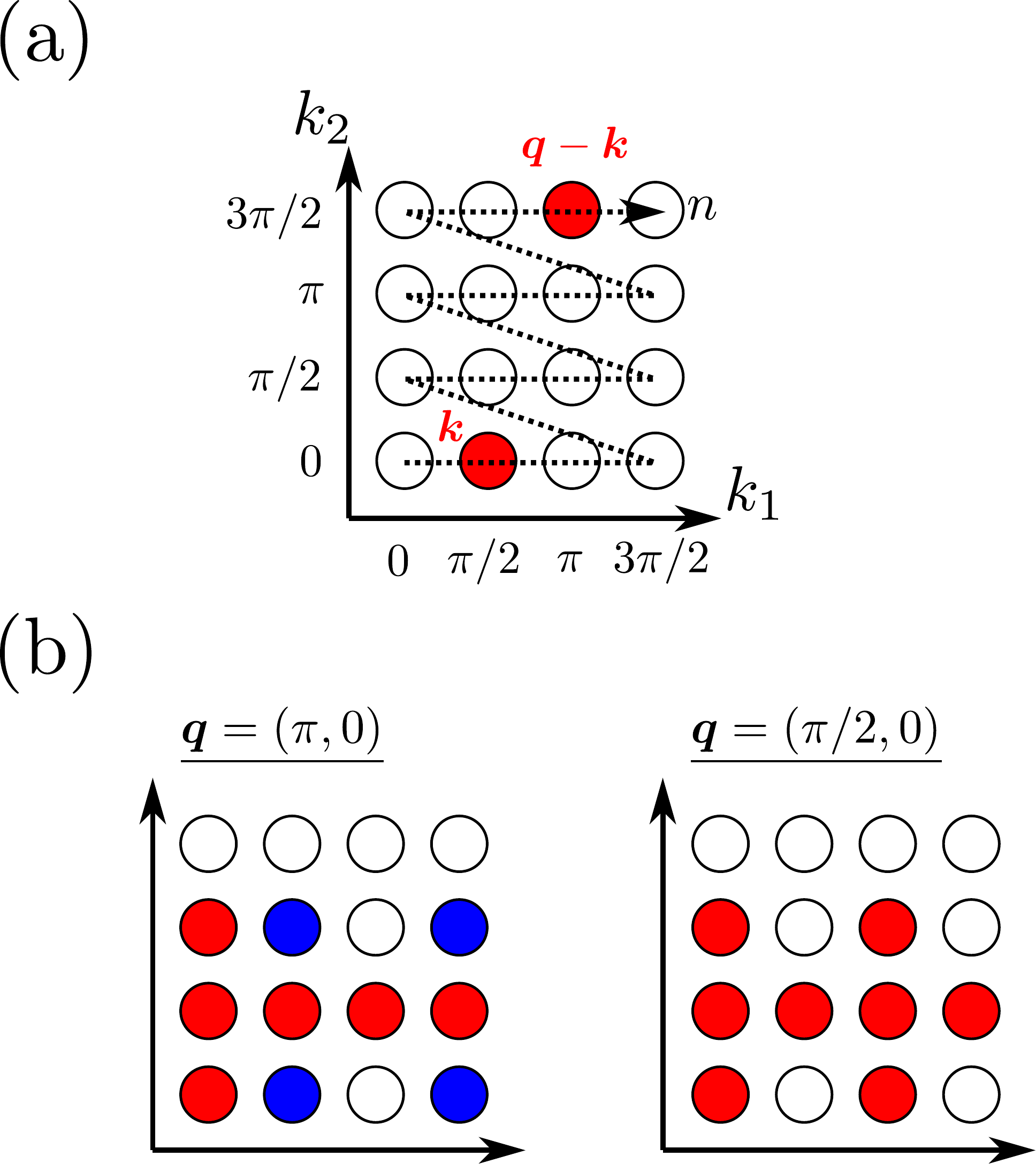}
 \caption{Two-particle configuration in momentum space [$L_1=L_2=4,~\bm{G}_1=(2\pi,0),~\bm{G}_2=(0,2\pi)$]. (a) Example of $(\bm{k},\bm{q}-\bm{k})$ that satisfies $n(\bm{k})\leq n(\bm{q}-\bm{k})$.  (b) Examples of momentum region $\{\bm{k}~|~n(\bm{k})\leq n(\bm{q}-\bm{k})\}$ for given $\bm{q}$. At the blue dots, $\bm{k}\equiv\bm{q}-\bm{k}~(\mathrm{Mod}~\bm{G}_{1,2})$. } 
 \label{fig1}
\end{center}
\end{figure}

In this paper, we focus on the energy spectrum of two-dimensional lattice systems with only two bosons.
Such a two-particle problem in bosonic/fermionic systems is widely formulated in various contexts. For example, see Refs. \onlinecite{Valiente-Petrosyan-08, Salerno-18, Iskin2021}.
We here derive the matrix representation of the two-particle Hamiltonian.
The Hilbert space of the two-particle system is spanned by the following basis vectors with the total momentum of two particles, $\bm{q}$:
\begin{align}
    &|\bm{q},\bm{k}\rangle:=\notag\\
    &\begin{cases}
        c^{\dagger}_{\bm{q}-\bm{k},\alpha}c^{\dagger}_{\bm{k},\alpha}|0\rangle&\mathrm{for~}\bm{k}\not\equiv\bm{q}-\bm{k}~(\mathrm{Mod}~\bm{G}_{1,2})\\
        c^{\dagger}_{\bm{q}-\bm{k},\alpha}c^{\dagger}_{\bm{k},\alpha}|0\rangle/\sqrt{2}&\mathrm{for~}\bm{k}\equiv\bm{q}-\bm{k}~(\mathrm{Mod}~\bm{G}_{1,2})
    \end{cases}.
\end{align}
Here the factor $\sqrt{2}$ comes from the bosonic nature.
To avoid the redundancy of the basis vectors, we have labeled the two-dimensional momenta $\bm{k}$ by integer numbers $n(\bm{k}):=n_1+L_1n_2$ and consider the cases with $n(\bm{k})\leq n(\bm{q}-\bm{k})$ [Fig.\ref{fig1}(a)]. 
By using the discrete translation invariance, $\bra{\bm{q},\bm{k}}H_{\rm proj}\ket{\bm{q}',\bm{k}'}\propto\delta_{\bm{q},\bm{q}'}$, we obtain the two-particle Hamiltonian:
\begin{align}
    H_{2p}&=\sum_{\bm{q}}{\sum_{\bm{k}}}'{\sum_{\bm{k}'}}' \ket{\bm{q},\bm{k}}\bra{\bm{q},\bm{k}}H_{\rm proj}\ket{\bm{q},\bm{k}'}\bra{\bm{q},\bm{k}'}\notag\\
    &=:\sum_{\bm{q}}{\sum_{\bm{k}}}'{\sum_{\bm{k}'}}'\mathcal{H}^{\bm{q}}_{\bm{k},\bm{k}'}|\bm{q},\bm{k}\rangle\langle\bm{q},\bm{k}'|.
\end{align}
The energy spectrum is given by the eigenspectrum of the $\bm{q}$-dependent Hamiltonian matrix $\mathcal{H}^{\bm{q}}$. Note that the summation ${\sum_{\bm{k}}}'$ does not span the whole Brillouin zone, and the momentum region in summation depends on the total momentum $\bm{q}$ [Fig.\ref{fig1}(b)].

A key step to understand the nature of $\mathcal{H}^{\bm{q}}$ is to introduce a matrix $A_{\bm{q}}$ whose elements are given by
\begin{align}
    &[A_{\bm{q}}]_{i,\bm{k}}=\notag\\
    &
    \begin{cases}
    \sqrt{\frac{4U}{L_xL_y}}u_{\bm{q}-\bm{k},\alpha}(i)u_{\bm{k},\alpha}(i)&\mathrm{for~}\bm{k}\not\equiv\bm{q}-\bm{k}\\
    \sqrt{\frac{2U}{L_xL_y}}u_{\bm{q}-\bm{k},\alpha}(i)u_{\bm{k},\alpha}(i)&\mathrm{for~}\bm{k}\equiv\bm{q}-\bm{k}
    \end{cases}.\label{non-square}
\end{align}
Importantly, the explicit form of $\mathcal{H}^{\bm{q}}$ is given in terms of $A_{\bm{q}}$:
\begin{align}
    \mathcal{H}^{\bm{q}}=A^{\dagger}_{\bm{q}}A_{\bm{q}}.\label{largeham}
\end{align}
As indicated from Eq. (\ref{largeham}), $\mathcal{H}^{\bm{q}}$ is a positive semi-definite matrix, and its eigenvalues are nothing but the square of singular values of $A_{\bm{q}}$.
For a large system size, the number of the columns of $A_{\bm{q}}$ $\sim\mathcal{O}(N_{\rm unit})$ is much larger than that of the number of rows that is equal to the internal degrees of freedom, $n_{\rm in}$.
Thus, the number of finite eigenenergies at each total momentum is at most $n_{\rm in}$, and the other majority of eigenstates become exact zero modes.
The former and latter modes are called ``bound states" and ``scattering (continuum) states", respectively \cite{Valiente-Petrosyan-08}. In the scattering states, two particles are not bounded and move around independently.
In the bound states, the two particles are bounded and feel the on-site Hubbard interaction.
When the kinetic part $H_0$ is included, exact zero-energy bands (continuum bands) acquire finite energy.

Although the two-particle spectrum itself can be calculated by the above formalism, $\mathcal{H}^{\bm{q}}$ is not appropriate for investigating band topology of the two-particle states.
As we noted, the region in which $\bm{k}$ runs depends on $\bm{q}$, and so does the basis set $\{\ket{\bm{q},\bm{k}}\}$. For the same reason, the dimension of $\mathcal{H}^{\bm{q}}$ can vary with respect to $\bm{q}$. 
Moreover, $\mathcal{H}^{\bm{q}}$ is changed under the $U(1)$ gauge transformation of the Bloch wave function:
\begin{align}
    \bm{u}_{\bm{k},\alpha}\rightarrow e^{i\chi(\bm{k})}\bm{u}_{\bm{k},\alpha}.\label{gauge}
\end{align}
Thus, $\mathcal{H}^{\bm{q}}$ is essentially discontinuous and not useful in topological characterization.
Then, the key idea of the present work is that,
instead of $\mathcal{H}^{\bm{q}}$, we focus on the $n_{\rm in}\times n_{\rm in}$ reduced matrix:
\begin{align}
    h^{\bm{q}}=A_{\bm{q}}A^{\dagger}_{\bm{q}}.\label{smallham}
\end{align}
Since $\mathcal{H}^{\bm{q}}$ and $h^{\bm{q}}$ share the finite eigenvalues, the two-particle Hamiltonian is formally given by
\begin{align}
    H_{2p}=\sum_{\bm{q}}\sum_{i,j}h^{\bm{q}}_{i,j}|\bm{q},i\rangle\langle\bm{q},j|.
\end{align}
Remarkably, $h^{\bm{q}}$ is invariant under the $U(1)$ gauge transformation (\ref{gauge}) and is continuous with respect to $\bm{q}$.
Roughly speaking, the index $i$ represents the orbital on which two bound particles are placed.

\section{Two-particle topology and fractional Chern insulator\label{two-particle-topology}}
\subsection{Relationship between two-particle spectrum and FCI (review)}
In addition to the one-particle topological/geometrical properties, the two-particle band spectrum is also used to predict the FCI ground states \cite{Lauchli-Liu-Bergholtz-Moessner-13,Liu-Bergholtz-Kapit-13}.
We here briefly review this idea.

We begin with addressing the FQHE, before considering the FCI.
Let us consider the two-particle problem in continuous space under a uniform magnetic field.
In the first quantization, a typical Hamiltonian is given in the following form \cite{yoshioka-textbook-02}:
\begin{align}
    H_{2p}=&\left[\frac{1}{2\mu}\left\{\bm{p}-\frac{e}{2}\bm{A}(\bm{r})\right\}^2+V(|\bm{r}|)\right]\notag\\
    &+\frac{1}{2M}\left\{\bm{P}-2e\bm{A}(\bm{R})\right\}^2,
\end{align}
where $\bm{A}$ represents the vector potential for a uniform magnetic field, $V$ represents the rotation-symmetric interaction, $\bm{r}$/$\bm{R}$ is the relative/center-of-mass coordinate, $\bm{p}$/$\bm{P}$ is the relative/center-of-mass momentum, $\mu$/$M$ is the reduced/total mass, and $e$ is the elementary charge.
As the Hamiltonian is divided into relative and center-of-mass parts, the eigenstates are given in the form $\phi(\bm{r})\Phi(\bm{R})$.
Remarkably, the center-of-mass part of the Hamiltonian has the same form as the Hamiltonian of the one-particle Landau problem, while the relative part has an additional term that divides the degeneracy of the Landau level into relative angular-momentum sectors.
When we focus on the lowest Landau level under a strong magnetic field, the relevant Hamiltonian has the following diagonal form:
\begin{align}
    H_{2p}=\sum_m\sum_\mathcal{M} V_m \ket{m,\mathcal{M}}\bra{m,\mathcal{M}},\label{Landau-two-particle}
\end{align}
where $m$/$\mathcal{M}$ is relative/center-of-mass angular momentum, and the relevant Hilbert space consists of two-particle states $\{\ket{m,\mathcal{M}}\}$. Thus, each two-particle ``band" is flat and characterized by $m$.
For bosonic/fermionic particles, $V_m$ can take finite value only for even/odd $m$ because of the symmetry/anti-symmetry.  
Note that the two-particle energy $V_m$ plays an important role in the FQH system.
Actually, the bosonic/fermionic Laughlin states at $\nu=1/q$ are the zero-energy eigenstates of the projected interaction Hamiltonian with $V_m>0,m\leq q-2$ and $V_m=0,m>q-2$ for even/odd $m$. In this context, the two-particle energy $V_m$ is known as Haldane's pseudopotential \cite{Haldane-Hierarchy-83}. 
Despite the absence of rotation symmetry, a similar construction can be performed for any geometry including the torus.

The idea of the characterization of many-body physics by the two-particle energy spectrum was imported into the area of the FCI, while the exact correspondence clearly does not hold. For example, the two-particle band structure in momentum space has finite bandwidth in general, unlike the flat ``band" in the two-particle Landau problem.
It was pointed out in Ref. \onlinecite{Lauchli-Liu-Bergholtz-Moessner-13} that each pseudopotential $V_m$ corresponds to a pair of two approximately-degenerate two-particle bands in the Brillouin zone.
If there are two approximately-degenerate bands whose energies are much larger than those of the other bands, it roughly indicates the stability of the $\nu=1/2$ bosonic FCI state.
Because of the lack of exact correspondence, the pair of energy eigenvalues are split significantly in general, while such a split is suppressed in the Kapit-Mueller (KM) model with a large magnetic cell \cite{Liu-Bergholtz-Kapit-13},
whose exact ground state is described by the lattice analog of the Laughlin state.
Note that if there are no internal degrees of freedom ($u_{\bm{k}}=1$), the two-particle band structure with finite eigenvalues consists only of one flat band.
In other words, the complicated multi-band nature of the two-particle spectrum is induced by the interband effect (e.g., topological and geometrical effects) between the projected band and the other bands in the one-particle spectrum.

\subsection{Similarity between center-of-mass Landau level and two-particle Chern band}
Although the many-body physics may not be explained completely only by the one-particle and two-particle properties, they sometimes provide guidelines for the FCI search.
There should remain other perspectives in this direction.
In this paper, we focus on the topological nature of the two-particle band structure.

To gain an insight, let us again consider the two-particle Hamiltonian (\ref{Landau-two-particle}) in the Landau problem.
Once we specify the relative angular momentum $m$, the only remaining degree of freedom is the center-of-mass angular momentum. 
Thus, for each $m$, the relevant Hilbert space consists of the wave functions in the lowest Landau level defined for the center-of-mass coordinate.
By focusing on the similarity between the Landau level and the Chern band, we here propose that a nontrivial Chern number of the finite-energy two-particle band structure in center-of-mass (total) momentum space roughly indicates the presence of the FCI ground states.
As we mentioned, the reduced matrix $h^{\bm{q}}$ is a useful matrix representation of the two-particle Hamiltonian for topological characterization.
Because the finite-energy states of $h^{\bm{q}}$ can be regarded as bound states of two particles, 
the two-particle topology is equivalent to the topology of the band structure of bound states.
In the reduced Hamiltonian, the degree of freedom about the relative coordinate is traced out.
This is similar to the two-particle Landau problem, in which the relative angular momentum is frozen for a fixed pseudopotential $V_m$ [see Eq. (\ref{Landau-two-particle})].

\begin{figure}[]
\begin{center}
 \includegraphics[width=8cm,angle=0,clip]{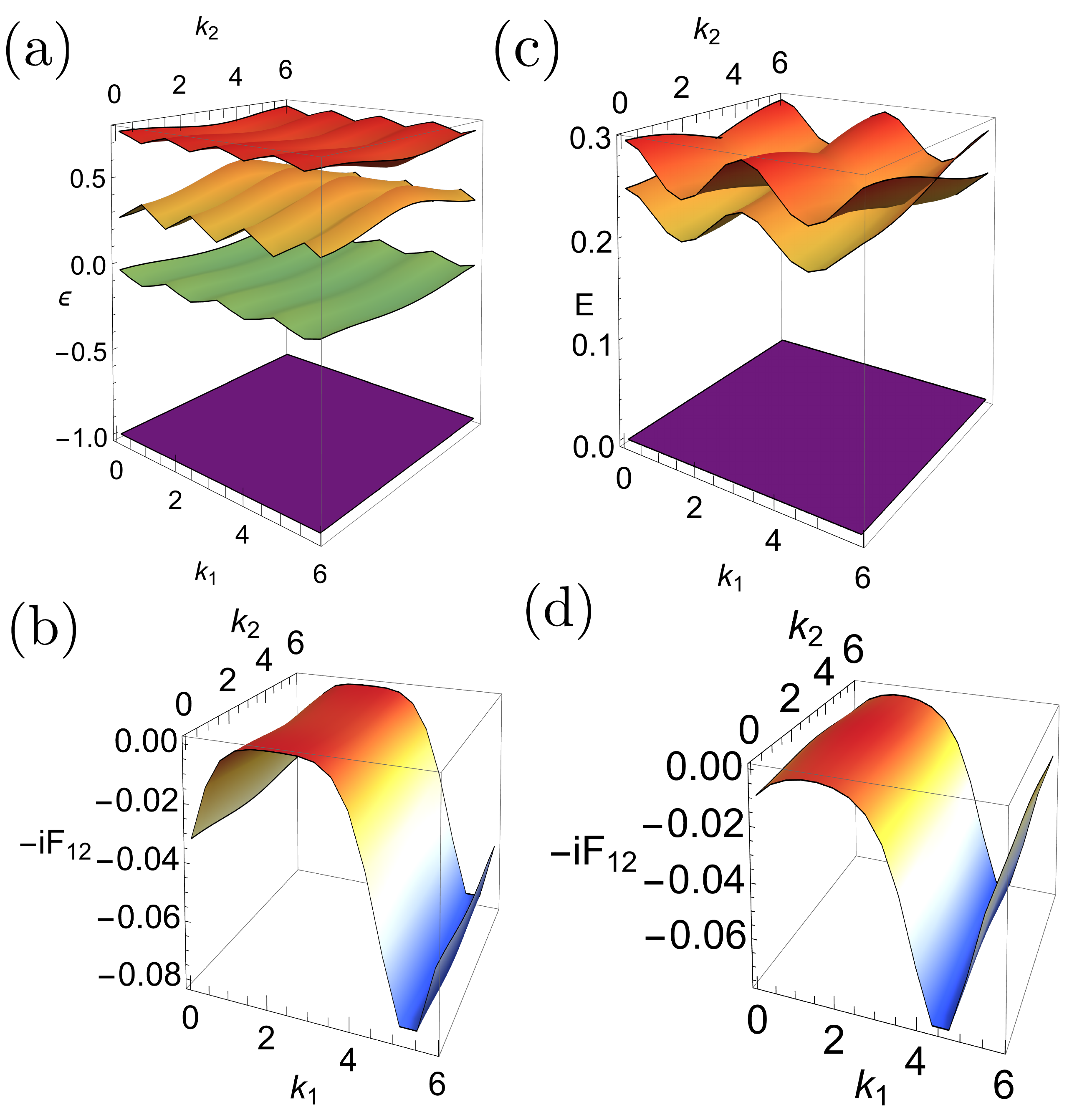}
 \caption{One-particle and two-particle properties of the KM model (PBC, Landau gauge, $L_1=L_2=16$). (a, c) One-particle and two-particle band structures. (b, d) 
 One-particle and two-particle Berry curvature in the momentum space.} 
 \label{fig2}
\end{center}
\end{figure}

As an example, we investigate the two-particle topology of the KM model~\cite{Kapit-Mueller-10}.
The KM model is a two-dimensional tight-binding model that mimics the Landau level, and the number of bands is equal to the number of sites in a magnetic unit cell, $q$, under the Landau gauge. The details of the model are given in Appendix~\ref{app:KM model}.
The lowest band of the KM model is an exactly flat band with Chern number $|C|=1$ owing to the infinite hopping range.
In order to reduce the numerical cost, we truncate the KM model to third neighbor hoppings and consider the four-band case $(q=4)$ with $16\times16$ magnetic unit cells.
The one-particle energy dispersion is shown in Fig.\ref{fig2} (a).
Despite the truncation, the lowest band is almost flat, which indicates that the essence of the KM model is not broken by the truncation.
The Chern number is calculated by the integral of the Berry curvature; See Appendix~\ref{app:Fukui-Hatsugai} for the details of the numerical method~\cite{Fukui-Hatsugai-Suzuki-05}.
The numerical integration shows that the Chern number of the lowest band is given by $C= -1$.
We calculate the two-particle band structure by the diagonalization of the $4\times4$ matrix $h^{\bm{q}}$ or the singular value decomposition of the $4\times16^2$ matrix $A^{\bm{q}}$ [Fig. \ref{fig2} (c)].
There are two non-separable finite-energy bands, and the other bands are exact zero-energy flat bands.
The total Berry curvature of the finite-energy eigenstates of $h^{\bm{q}}$ is shown in Fig. \ref{fig2}(d).
As expected, the numerical integration indicates the nontrivial Chern number ($C=-1$). 
In Sec. \ref{many-body-calculation}, we numerically check that the ground states of the truncated KM model at $\nu=1/2$ are the FCI states, which supports our proposal about the correspondence between the two-particle topology and the FCI ground states.

\subsection{Two-particle topology and bulk-boundary correspondence}
In this subsection, we consider the meaning of the two-particle topology of $h^{\bm{q}}$ in terms of the bulk-boundary correspondence.
Before proceeding to detailed discussions, it is worth noting the following fact: 
In the case of one-particle Hamiltonian matrix $H(\bm{k})$, the internal degree $i$ and momentum $\bm{k}$ are physically independent of each other, and one can construct the real-space OBC Hamiltonian by performing the inverse Fourier transform of $H(\bm{k})$. 
If we use another basis for a momentum-dependent matrix instead of the internal degree $i$, the matrix elements of the real-space OBC Hamiltonian are not given by the inverse Fourier transform in general. 
For example, if we work on the band basis, band topology with respect to this basis is trivial, and the OBC matrix constructed from the inverse Fourier transform has no topological edge states.
In this sense, the physical meaning of the basis is important when we consider band topology and bulk-boundary correspondence.

The above discussion indicates that one should carefully interpret the two-particle topology of $h^{\bm{q}}$.
Strictly speaking, the inverse Fourier transform of $h^{\bm{q}}$ does not give the matrix elements of the OBC Hamiltonian because of the definition of $h^{\bm{q}}$, Eqs. (\ref{non-square}) and (\ref{smallham}), contains the one-particle wave functions specific to the PBC. 
Thus, the index $i$ and momentum $\bm{q}$ are not physically independent of each other, unlike the one-particle case. Nevertheless, one can still find the bulk-boundary correspondence for the two-particle topology. In the following, we formulate the two-particle problem under the OBC and discuss the bulk-boundary correspondence numerically.

\begin{figure}[]
\begin{center}
 \includegraphics[width=0.99\linewidth,angle=0,clip]{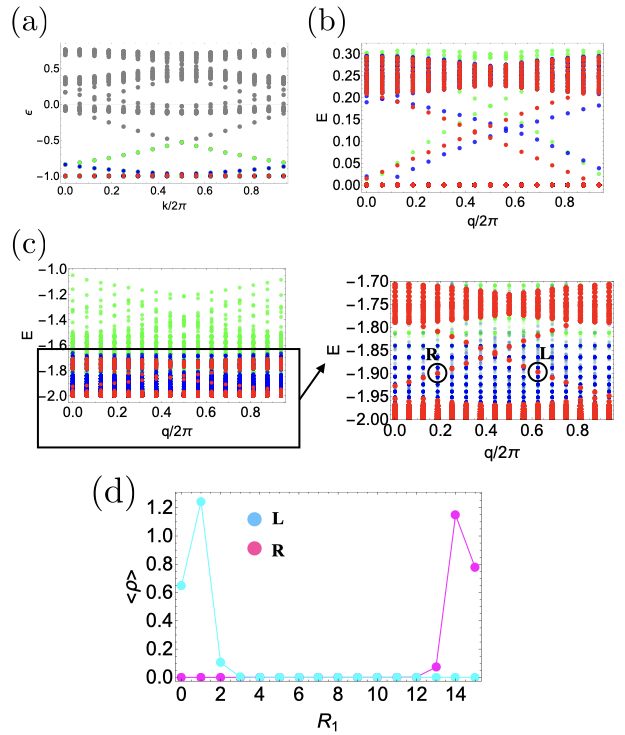}
 \caption{(a) One-particle spectrum of the KM model under the OBC. 
 Gray dots represent the full spectrum. 
 Green, blue, and red dots represent the $n_{\rm proj}$ lowest states with $n_{\rm proj} = L_1$, $L_1-1$, and $L_1-2$, respectively.
 (b,c) The two-particle spectrum of the KM model under the OBC. 
 For (b), the single-particle energy is neglected [Eq.~(\ref{twoparticleobcwo})], whereas it is included for (c) [Eq.~(\ref{eq:twoparticle_ham_obc_full})]. The colors of the dots denote the different choices of $n_{\rm proj}$ in the same way as (a). 
 (d) Real-space particle density of the left (L) and right (R) edge modes,
 indicated by the black circles in panel (c). We have set $L_1=L_2=16$.
 } 
 \label{fig3}
\end{center}
\end{figure}

We impose the OBC in the $x$ direction and the PBC in the $y$ direction.
We repeat almost the same procedure discussed in Sec. \ref{formalism}.
Instead of two-dimensional momentum, we introduce one-dimensional momentum in the $y$ direction $k=n_2G/L_2$ with $G$ being the one-dimensional reciprocal lattice vector: 
\begin{align}
    c_{\bm{R},i}=\frac{1}{\sqrt{L_2}}\sum_{k}\sum_{a}u_{k,a}(R_1,i)e^{ikR_2}c_{k,a}. \label{eq:Fourier_OBC}
\end{align}
The interaction Hamiltonian is given by
\begin{align}
    &H_{\rm int}=\frac{U}{L_2}\sum_{q,k,k'}\sum_{R_1,i}\sum_{a,b,a',b'}\notag\\
    &u^{*}_{k,a}(R_1,i)u^{*}_{q-k,b}(R_1,i)u_{q-k',b'}(R_1,i)u_{k',a'}(R_1,i)\notag\\
    &c^{\dagger}_{k,a}c^{\dagger}_{q-k,b}c_{q-k',b'}c_{k',a'}. \label{eq:intHam_OBC}
\end{align}
In Eq.~(\ref{eq:intHam_OBC}), 
for the summation over $a$, $a^\prime$, $b$, and $b^\prime$ which label the one-particle eigenstate at each $k$, 
we have to set the upper limit, $n_{\rm proj}$,
to make the correspondence to the PBC problem 
where we consider the interaction Hamiltonian projected onto the lowest band. 
Specifically,
$n_{\rm proj}$ should be $\mathcal{O}(L_1)$ at each one-dimensional momentum.
The Hilbert space of the two-particle system is spanned by the following basis vectors with total one-dimensional momentum of two particles:
\begin{align}
    &|q;k,ab\rangle:=\notag\\
    &\begin{cases}
        c^{\dagger}_{q-k,b}c^{\dagger}_{k,a}|0\rangle&\mathrm{for~}k\not\equiv q-k~(\mathrm{Mod}~G)~\mathrm{or}~a\neq b\\
        c^{\dagger}_{q-k,b}c^{\dagger}_{k,a}|0\rangle/\sqrt{2}&\mathrm{for~}k\equiv q-k~(\mathrm{Mod}~G)~\mathrm{and}~a=b
    \end{cases}. \label{eq:twoparticle_obc_basis}
\end{align}
As in the case of full PBC, we label $(k,a)$ by integer number $n(a,k)=a+n_{\rm proj}n_2$ and consider the cases with $n(a,k)\leq n(b,q-k)$. 
The two-particle Hamiltonian under the present boundary condition is given by
\begin{align}
H_{2p}&=\sum_{q}{\sum_{kab}}'{\sum_{k'a'b'}}'\mathcal{H}^{q}_{kab,k'a'b'}|q;k,ab\rangle\langle q;k',a'b'|\notag\\
    &=\sum_{q}\sum_{R_1i,R_1'i'}h^{q}_{R_1i,R_1'i'}|q,R_1i\rangle\langle q,R_1'i'|,
\end{align}
where     
\begin{align}
&\mathcal{H}^{q}=A^{\dagger}_{q}A_{q},~h^{q}=A_{q}A^{\dagger}_{q},\notag\\
    &[A_{q}]_{R_1i,kab}
    =\notag\\
    &\begin{cases}
    \sqrt{\frac{4U}{L_y}}u_{q-k,b}(R_1,i)u_{k,a}(R_1,i)&\mathrm{for~}k\not\equiv q-k~\mathrm{or}~a\neq b\\
    \sqrt{\frac{2U}{L_y}}u_{q-k,b}(R_1,i)u_{k,a}(R_1,i)&\mathrm{for~}k\equiv q-k~\mathrm{and}~a=b
    \end{cases}.\label{twoparticleobcwo}
\end{align}
If we choose the PBC instead of the OBC, the band index $a$ is interpreted as crystal momentum in the $x$ direction under the projection onto the lowest band. 

Now we are in a position to investigate the bulk-boundary correspondence that originates from the two-particle topology.
In the actual construction of the OBC two-particle Hamiltonian, however, the choice of $n_{\rm proj}$ is not so obvious because of the presence of gapless boundary states that connect the gapped bulk bands in the one-particle spectrum [Fig.\ref{fig3}(a)].
In addition,
as the one-particle energies of the gapless boundary states are not negligible, 
we also calculate the two-particle spectrum by diagonalizing a matrix with the following elements:
\begin{align}
    &\mathcal{H}^q_{kab,k'a'b'}=\notag\\&[A^{\dagger}_{q}A_{q}]_{kab,k'a'b'}+\delta_{k,k'}\delta_{a,a'}\delta_{b,b'}(\epsilon_{k,a}+\epsilon_{q-k,b}), \label{eq:twoparticle_ham_obc_full}
\end{align}
where $\epsilon_{k,a}$ is the one-particle energy under the OBC.
Note that one can no longer construct the reduced matrix $h^{q}$ in the presence of the one-particle energy.

By using the expressions (\ref{twoparticleobcwo}) and (\ref{eq:twoparticle_ham_obc_full}), we numerically investigate the two-particle bulk-boundary correspondence of the KM model with/without one-particle energies for various $n_{\rm proj}$.
Under the full PBC, the dominant bands of the two-particle spectrum have unit Chern number $(C=-1)$, as calculated in the previous section.
First, let us consider the two-particle spectrum in the absence of one-particle energies.
In this case, there exist gapless edge states that connect the bulk bands for various $n_{\rm proj}$, while the details of the spectra depend on $n_{\rm proj}$ [Fig.\ref{fig3}(b)].
These gapless edge states can be regarded as a consequence of the nontrivial Chern number ($C = -1$) of two-particle bands. 
In the presence of one-particle energies, however, the two-particle spectra have more complicated structures.
For $n_{\rm proj}=L_1$, not only the states in the bulk flat band but also those in the edge states with dispersion are included in the projected states.
If scattering states contain edge states in the one-particle spectrum, they can also have energy dispersion. 
Although such states are essentially different from the edge states of two-particle bound states, they are also found in the two-particle spectrum [Fig. \ref{fig3}(c)]. Thus, the scattering states and edge states from two-particle bound states coexist at the edges for $n_{\rm proj}=L_1$.
For smaller $n_{\rm proj}$, dispersion of the scattering states is suppressed, and the two-particle spectrum looks similar to that in the absence of the one-particle energies [Fig. \ref{fig3}(c)].
Again, the remaining edge states can be regarded as a consequence of the two-particle topology.
The real-space particle density of such edge states are plotted in Fig. \ref{fig3}(d) (see Appendix~\ref{app:density} for details of the calculation).
If the one-particle band gap divided by $L_2$, which approximates the level distance between the edge states, is much larger than the interaction $U$, the projection only onto the bulk flat band is physically reasonable.
Our numerical calculations indicate that the bulk-boundary correspondence holds for the two-particle Chern number.
Thus, the matrix elements that are calculated by the inverse Fourier transform of the $h^{\bm{q}}$ in the $x$ direction may well approximate those of the true two-particle Hamiltonian under the present boundary condition, $h^{q}$, if the scattering edge states are negligible.

\begin{figure}[]
\begin{center}
 \includegraphics[width=6cm,angle=0,clip]{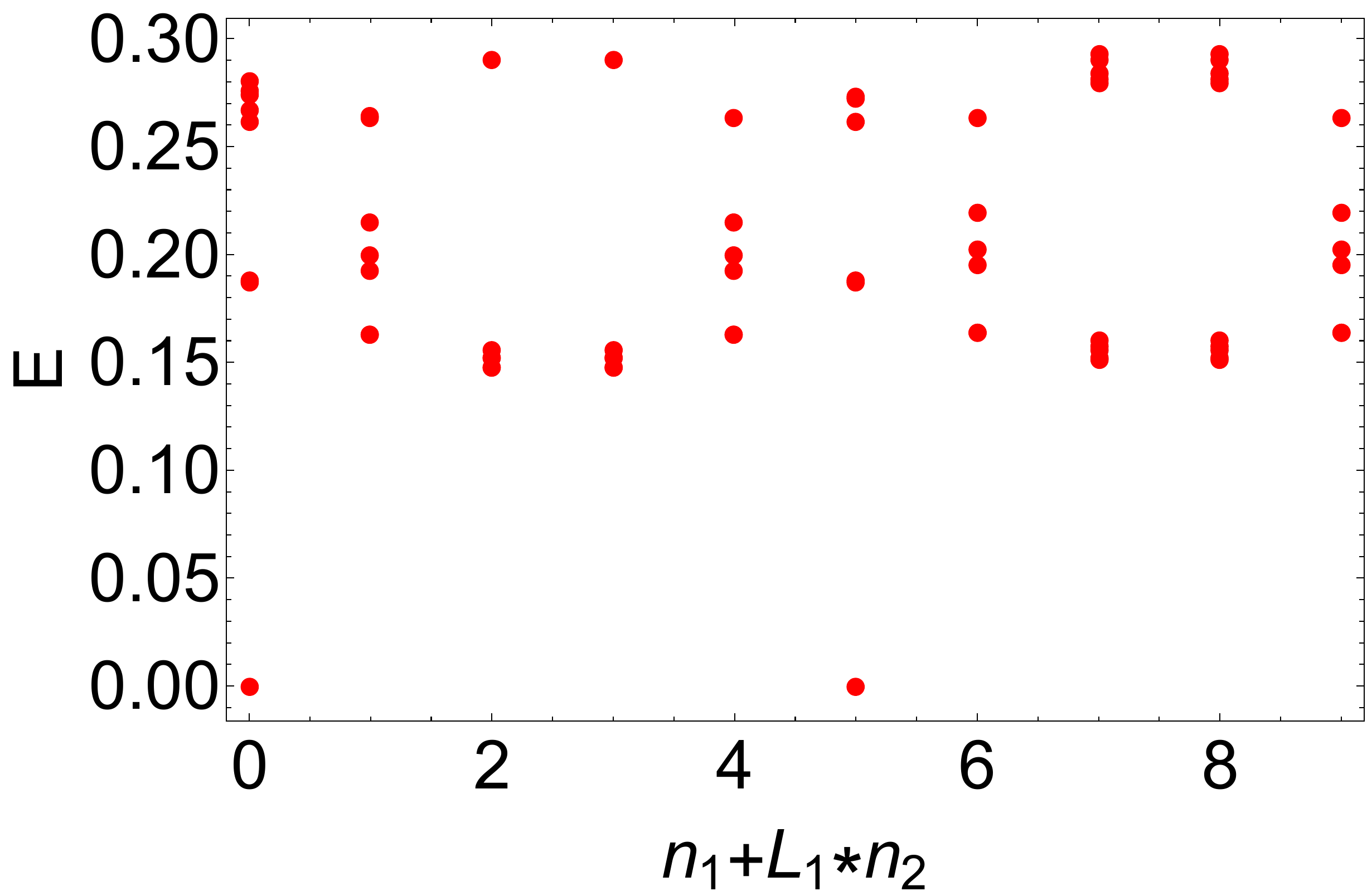}
 \caption{Many-body eigenspectrum of projected Hamiltonian for truncated KM model (PBC, $L_1=5,L_2=2$, $q=4$). In each momentum sector labeled by $(n_1,n_2)$, the eigenvalues are plotted from smallest to sixth. } 
 \label{fig4}
\end{center}
\end{figure}
\subsection{Many-body ground state at $\nu = 1/2$\label{many-body-calculation}}
Reference \onlinecite{Kapit-Mueller-10} showed that the $\nu=1/2$ bosonic Laughlin wave function \cite{Kalmeyer-Laughlin-87} is an exact many-body ground state of the 
(untruncated) KM model in the presence of local interactions at $\nu=1/2$. Owing to the topological nature, the ground states under the PBC have two-fold topological degeneracy at $\nu=1/2$ in infinite-volume limit.
We here numerically check that the ground states of the many-body Hamiltonian $H_{\rm proj}$ for the truncated KM model ($q=4$ and including up to the third neighbor hoppings) in the previous subsections are in the FCI phase at $\nu=1/2$.
We perform the numerical calculations by
using the code based on that of
Ref.~\onlinecite{Varjas-Abouelkomsan-Kang-Bergholtz-22}.
%
Thanks to the discrete translation invariance, the total momentum is a good quantum number under the PBC.
In Fig. \ref{fig4}, the eigenvalues are plotted from smallest to sixth in each momentum sector.
As expected, the first and second lowest energy eigenvalues are degenerated within numerical accuracy and separated from the other states with a large gap, while the ground-state degeneracy is not exact in a usual FCI model at a finite system size.
We regard these almost degenerated states as the ground states henceforth.
The momentum sectors in which the ground states exist can depend on the system size.
In the present calculation, the two ground states are placed in different sectors, $(n_1,n_2)=(0,0),(0,1)$.
In the FCI phase, the ground states should not mix with the excited states under magnetic flux insertion through the handle of the torus \cite{Bergholtz-Liu-13}.
Typically, the two ground-state energy eigenvalues are reversed after $2\pi$-flux insertion, and it takes $4\pi$ flux in total to recover the original spectrum.

In the present model, the Chern number of the dominant two two-particle bands is nontrivial ($C=-1$). At the same time, the many-body spectrum indicates the presence of topological degeneracy at $\nu=1/2$. In this case, the two-particle topology seems to be related to the stability of the FCI ground states.
Then the following question naturally arises: does this correspondence universally hold in FCI models?

\begin{figure*}[]
\begin{center}
 \includegraphics[width=0.95\linewidth,angle=0,clip]{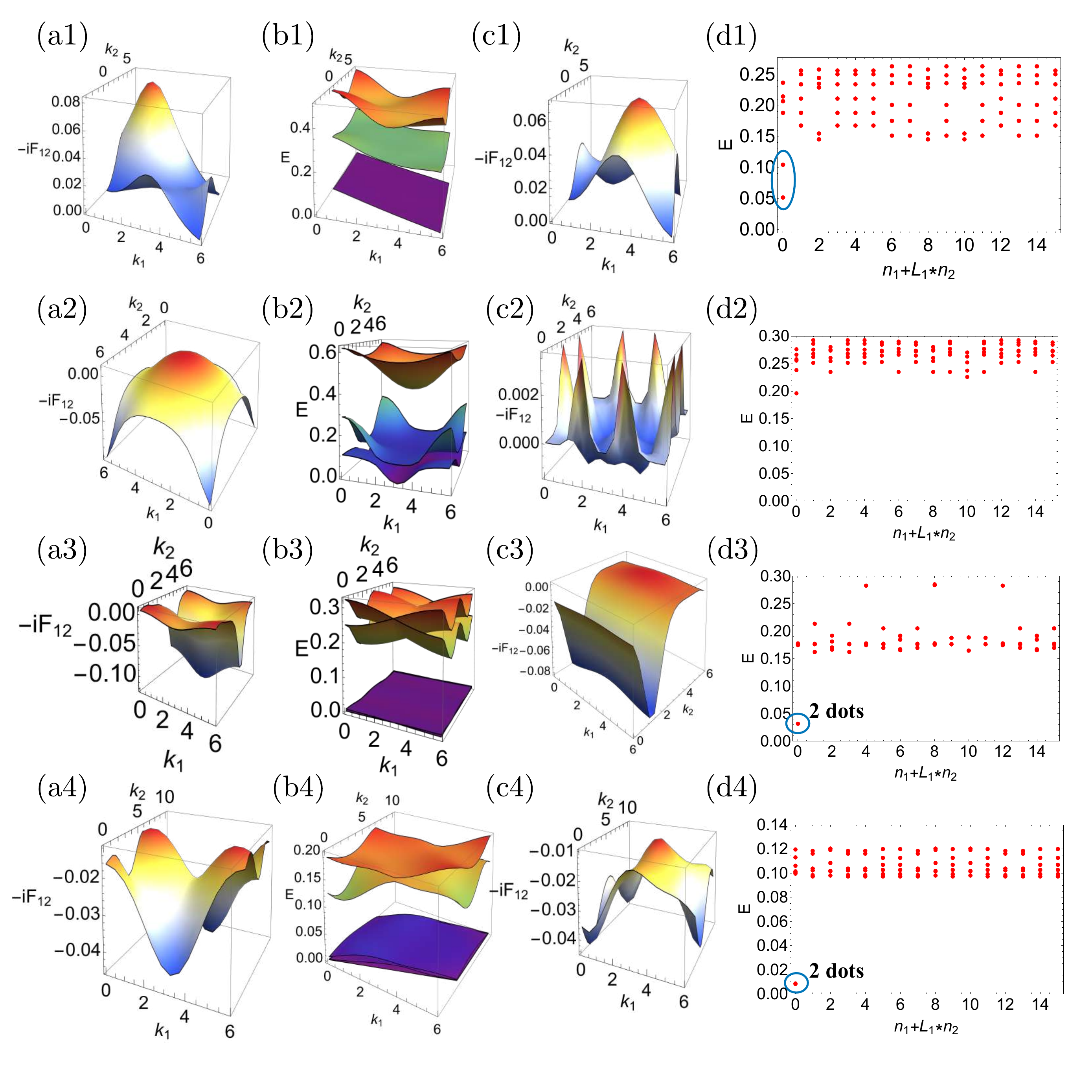}
 \caption{One-, two-, and many-particle properties of $n_{\rm in}\geq3$-band models. Results for Kagome, square, Hofstadter, and ruby lattice models are shown from top to bottom. See Appendix~\ref{app:tbmodels} for details of models. (a) Momentum-space distribution of one-particle Berry curvature of the lowest band. (b) Two-particle band structure.  (c) Momentum-space distribution of two-particle Berry curvature of the dominant bands. In (a-c), the system size is $16\times16$. (d) Many-body eigenspectrum of projected Hamiltonian. The system size is $4\times4$.
 Note that the lowest two points at $n_1+L_1 n_2 = 0$ for the panels (d3) and (d4) are almost degenerate, which we mark by the blue circles for clarity.
 } 
 \label{fig5}
\end{center}
\end{figure*}

\section{Two-particle topology of various FCI models \label{sec:various_model}}

\begin{table}[b]
  \caption{Relationship between two-particle topology and $\nu=1/2$ bosonic FCI ground states. Ground-state degeneracy of the $4\times4$ system is used to determine whether or not the ground states are the FCI. The degree of degeneracy in the Kagome model, represented by ``Yes", is not as good as that of the other FCI models, represented by Yes. }
  \label{table1}
  \centering
  \begin{tabular}{c|c|c|c}
    \hline
    Model & $n_{\rm in}$& FCI & Two-particle topology\\
    \hline \hline
    Kagome~\cite{Tang-Mei-Wen-11} & 3 & ``Yes" & $C=1$ (upper two bands)\\
    square~\cite{Sun-Gu-Katsura-DasSarma-11} & 3 & No & trivial (one dominant band) \\
    KM~\cite{Kapit-Mueller-10} & 4 & Yes  & $C=-1$ (upper two bands) \\
    Hofstadter~\cite{Hofstadter-76} & 4 & Yes  & $C=-1$ (upper two bands) \\
    ruby~\cite{Hu-Kargarian-Fiete-11,Wu-Bernevig-Regnault-12} & 6 & Yes  & $C=-1$ (upper two bands)\\
    \hline
    QWZ~\cite{Qi-Wu-Zhang-06} &2&No&trivial (one dominant band)\\
    modified QWZ&2&Yes& trivial (small-gap semimetal)\\
     Haldane~\cite{Haldane-88,Neupert-Santos-Chamon-Mudry-11}&2&Yes& gapless (Dirac points)\\
    checkerboard~\cite{Neupert-Santos-Chamon-Mudry-11}
    &2&Yes& gapless (nodal line) \\
    \hline
  \end{tabular}
\end{table}

In this section, we numerically investigate the relationship between the two-particle topology and the $\nu=1/2$ FCI ground states for various tight-binding models whose lowest band is the Chern band with $|C|=1$. See Appendix~\ref{app:tbmodels} for details of models. 
As a many-body Hamiltonian, we analyze the bosonic projected Hamiltonian (\ref{projectedham}).
Because the projection disregards the one-particle kinetic energy, we use only the information of one-particle eigenstates of the given tight-binding model.
Note that the presence of the FCI phase under the fermionic interaction does not ensure the bosonic FCI ground states in the present case.

\begin{figure}[]
\begin{center}
 \includegraphics[width=8cm,angle=0,clip]{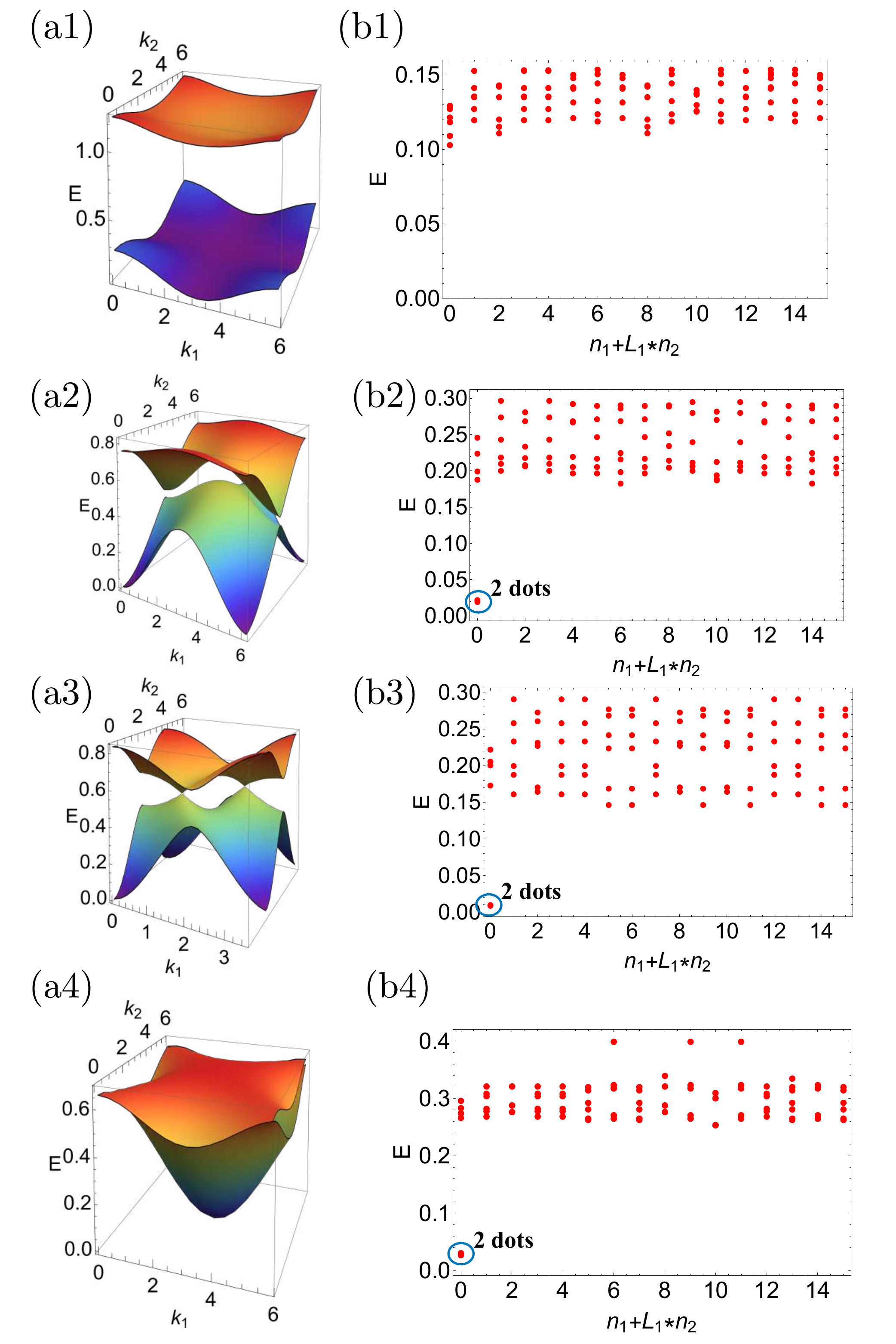}
 \caption{(a) Two-particle band structures and (b) many-body spectra of $n_{\rm in}=2$ models. Results for QWZ, modified QWZ, Haldane, and checkerboard lattice models are shown from top to bottom.
 See Appendix~\ref{app:tbmodels} for details of models. The system size is $16\times 16$ for (a) and $4\times4$ for (b). 
 We again mark the almost-degenerate two lowest energy states by the blue circles in the panels (b2), (b3), and (b4).} 
 \label{fig6}
\end{center}
\end{figure}

The numerical results are summarized in Table \ref{table1} and Figs. \ref{fig5} and \ref{fig6}.
The ground-state degeneracy of the $4\times4$ system is used to determine whether or not the ground states are the FCI states.
Further numerical justification of the FCI is a remaining task. 
For $n_{\rm in}\geq3$, the two-particle Chern number of the dominant upper bands well describes the presence or absence of the FCI ground states at $\nu=1/2$ \cite{Lauchli-Liu-Bergholtz-Moessner-13,Liu-Bergholtz-Kapit-13}. In all of the examples with FCI ground states (Yes in Table \ref{table1}), the number of dominant two-particle bands is two, which is consistent with previous studies that insist that each pseudopotential corresponds to two two-particle bands in FCI models.
Note that the degree of topological degeneracy in the Kagome model, represented by ``Yes" in Table \ref{table1}, is not as good as that of the other FCI models.
In this model, although the two-particle topology of the dominant two bands is nontrivial, the energy separation between these bands is relaticely large. 
For an ideal FCI model, both the two-particle topology and the degree of degeneracy of two-particle bands seem to be important.

For $n_{\rm in}=2$, however, the total Chern number of the two two-particle bands should be zero.
In this sense, the $n_{\rm in}=2$ case is essentially different from the other $n_{\rm in}$'s.
In Fig. \ref{fig6}, we omit the plots of one- and two-particle Berry curvature.
In models with FCI ground states, two two-particle bands tend to be connected with each other.
In these examples, the two-particle spectra host the gapless or small-gap semimetallic structures.
These complicated structures can be obstacles for the trivial two-band structures that lead to the non-FCI ground states at $\nu=1/2$.
Indeed, the QWZ model has no FCI-like ground state at $\nu=1/2$, and its two-particle spectrum seems to have no specific structures, as in the case of trivial one-band models. 
In contrast, the modified QWZ model (Appendix \ref{app:tbmodels}) has FCI-like ground states.
The modified QWZ model is obtained by changing the basis of the original QWZ model.
This change preserves the momentum-space distributions of the Berry curvature and quantum metric of the one-particle Chern band. 
Although these two properties are considered to be related to FCI stability, they cannot fully describe FCI physics.
In the present case, the projected many-body Hamiltonian and the two-particle spectrum are not invariant under the change of the basis.
It is worth mentioning 
that the uniqueness of two-band models is also found
in another indicator of the FCI. 
Specifically, it was proved in Ref.~\onlinecite{Varjas-Abouelkomsan-Kang-Bergholtz-22} 
that the Berry curvature of the two-band models can not have a uniform distribution in the momentum space,
while the constant Berry curvature is believed to be a favorable condition for realizing the FCI.

Although the two-particle topology is neither a necessary nor a sufficient condition for the FCI ground states at $\nu=1/2$ as other indicators in previous studies, the numerical results indicate that the two-particle topology characterizes the degree of similarity to the FQH system.
In any case, the interband nature of the two-particle band structure may be a key ingredient for the stabilization of the FCI ground states.

\section{Summary and discussions\label{summary-and-discussions}}
In this paper, we have investigated the topology in two-particle problems.
By using the reduced matrix representation, we have focused on the Hamiltonian of bound states among the whole states.
We have defined the Chern number of bound-state band structure and numerically investigated the bulk-boundary correspondence in the two-particle spectrum.
The shape of the two-particle band structure has already been known as one of the indicators for the stabilization of the FCI ground states in the many-body problem.
As another indicator, we have noticed the two-particle topology.
By focusing on the similarity between the Landau level and the Chern band, we have proposed that a nontrivial Chern number of the two-particle band structure roughly indicates the presence of the FCI ground states. 
Although the many-body properties should never be determined completely by the few-body physics, 
we have found several FCI models in which the proposed relation holds.

Before closing the paper, we point out several future works we will address.
Because the analogy between the two-particle Landau level and two-particle Chern number is not specific to the bosonic systems, we can also consider the same physics under the fermionic interaction.
In our paper, the internal degrees of freedom of the two-particle bound states coincide with those of the one-particle states owing to the bosonic on-site interaction.
Under the fermionic interactions, however, the two particles never occupy the same orbital, and the internal degrees of freedom of two-particle bound states are different from those of the one-particle problem.
Nevertheless, one can still define the reduced Hamiltonian $h^{\bm{q}}$ whose internal degrees of freedom consists of local two-particle orbitals.
A typical candidate for the fermionic FCI is the bilayer graphene. Recent theoretical and experimental studies about the bilayer graphene with FCI ground states are summarized in a review paper \cite{Liu-Bergholtz-review-22}.

Another important issue is the relationship between the one-particle and two-particle properties.
As we mentioned in the introduction, previous studies discussed the importance of momentum-space distributions of the Berry curvature and quantum metric in one-particle band structures \cite{Parameswaran-Roy-Sondhi-12,Parameswaran-13,Roy-geometry-14,Jackson-Moller-Roy-15,Claassen-Lee-Thomale-Qi-Devereaux-15,Lee-Claassen-Thomale-17,Mera-Ozawa-21-2,Varjas-Abouelkomsan-Kang-Bergholtz-22}.
In this direction, the roles of the momentum-space distributions of the one-particle quantities in the two-particle topology may be interesting topics.
In addition, one-particle bands with the higher Chern number are out of the scope of the present paper because of the absence of correspondence with the Landau levels.
Typically, the two-particle band structure for such cases  is highly complicated \cite{udagawa2014correlations}.
Thus, an extension of our theory to the higher Chern number is also a remaining future work.

\acknowledgements
N.O. is supported by JSPS KAKENHI Grant No.~JP20K14373 and JST CREST Grant No.~JPMJCR19T2, Japan.
T.M. is supported by JSPS KAKENHI Grant No.~JP20K14371.

\appendix
\section{Kapit-Mueller model \label{app:KM model}}
In this appendix, we explain the Kapit-Mueller (KM) model.
The KM model was introduced as a lattice implementation of Landau levels \cite{Kapit-Mueller-10}. In contrast to the Hofstadter model, this model allows infinite-range hopping terms whose strength decays exponentially with respect to the hopping range.
Without truncation of long-range terms, the lowest band of this model is an exactly flat band.
The explicit form is given by
\begin{align}
&H=\sum_{j\neq k}J(z_j,z_k)c^{\dagger}_jc_{k},\notag\\
&J(z_j,z_k)=W(z)e^{(\pi/2)(z_jz^*-z^*_jz)\phi},\notag\\
&W(z)=(-1)^{x+y+xy}e^{-(\pi/2)(1-\phi)|z|^2},
\end{align}
where $z_j=x_j+iy_j$, $z=x+iy=z_k-z_j$, and $\phi=p/q$ with $p$ and $q$ being two relatively prime numbers.
This representation corresponds to the symmetric gauge.
In the actual calculations, we use the Landau gauge, which is more convenient for the Fourier transform.

\section{Numerical calculation of Berry curvature and Chern number \label{app:Fukui-Hatsugai}}
In this appendix, we review the outline of the numerical calculation of Berry curvature and Chern number, based on 
Ref.~\onlinecite{Fukui-Hatsugai-Suzuki-05}.

Let $\ket{u_{\bm{k},\alpha}}$ 
be the eigenstate of the band $\alpha$
at momentum $\bm{k}$, which is non-degenerate in the entire Brillouin zone.
Numerically, we obtain $\ket{u_{\bm{k},\alpha}}$ at discretized 
points of the Brillouin zone as in Eq.~(\ref{eq:k_discrite}).
Then, we introduce the following variable, defined on a link between neighboring discretized momenta, $\mathcal{U}_{\mu}(\bm{k})$ ($\mu = 1,2$):
\begin{align}
\mathcal{U}_{\mu}(\bm{k}):= \frac{\langle u_{\bm{k},\alpha} \ket{ u_{\bm{k}+\frac{1}{L_\mu}\bm{G}_\mu,\alpha}}}{|\langle u_{\bm{k},\alpha} \ket{ u_{\bm{k}+\frac{1}{L_\mu}\bm{G}_\mu,\alpha}} |} ,
\end{align}
Using $\mathcal{U}_{\mu}(\bm{k})$, approximate Berry 
flux at a small plaquette whose left-bottom corner is at $\bm{k}$ is given as $-iF_{12} (\bm{k})$, where
\begin{align}
F_{12} (\bm{k}) = \ln [\mathcal{U}_{1}(\bm{k})\mathcal{U}_{2}(\bm{k}+\frac{1}{L_1}\bm{G}_1)\mathcal{U}^\ast_{1}(\bm{k}+\frac{1}{L_2}\bm{G}_2)\mathcal{U}^\ast_{2}(\bm{k})].
\end{align}
Consequently, the Chern number is approximated by 
the sum of the Berry flux of the plaquettes: 
\begin{align}
C = \frac{1}{2\pi i} \sum_{\bm{k}}F_{12} (\bm{k}).
\end{align}

When several bands degenerate at certain points or regions in the Brillouin zone, one can calculate the total Chern number for a set of such bands, $\alpha_1, \cdots, \alpha_n$,  by changing the link variables as
\begin{align}
    \mathcal{U}_{\mu}(\bm{k}):= \frac{\mathrm{det} [\Psi^\dagger(\bm{k}) \Psi(\bm{k}+\frac{1}{L_\mu}\bm{G}_\mu)]
    }{|\mathrm{det} [\Psi^\dagger(\bm{k}) \Psi(\bm{k}+\frac{1}{L_\mu}\bm{G}_\mu)]|},
\end{align}
where 
\begin{align}
    \Psi(\bm{k}) = \begin{pmatrix}
    \ket{u_{\bm{k},\alpha_1}} & \cdots & \ket{u_{\bm{k},\alpha_n}}\\
    \end{pmatrix}.
\end{align}

\section{Particle density in real space \label{app:density}}
In this appendix, we elucidate how to calculate the particle density in real space shown in Fig.~\ref{fig3}(d).

The particle density operator as a function 
of $R_1$ and $i$ is given as
\begin{align}
\rho_{R_1,i} = \sum_{R_2} c^\dagger_{(R_1,R_2,i)} c_{(R_1,R_2,i)}. \label{eq:density_real_1}
\end{align}
Substituting Eq.~(\ref{eq:Fourier_OBC}) into Eq.~(\ref{eq:density_real_1}), we have 
\begin{align}
\rho_{R_1,i} = \sum_{k} \sum_{a,b}
\mathcal{N}_{k,ab}(R_1,i)
c^\dagger_{k,a} c_{k,b}, \label{eq:density_real_2}
\end{align}
where we have defined $\mathcal{N}_{k,ab}(R_1,i):= u^\ast_{k,a}(R_1,i) u_{k,b}(R_1,i)$.
Equation (\ref{eq:density_real_2}) indicates that
the operating $\rho_{R_1,i}$ does not change the total momentum.
Hence, when acing $\rho_{R_1,i}$ on the state with the total momentum $q$, the resulting state has the same momentum. 

We now consider the expectation value of $\rho_{R_1,i}$.
Let $\ket{\Psi_q}$ be a normalized eigenstate of $\mathcal{H}^q$ of Eq.~(\ref{eq:twoparticle_ham_obc_full}),
which can be expanded by basis of Eq.~(\ref{eq:twoparticle_obc_basis}):
\begin{eqnarray}
\ket{\Psi_q} = \sum_{k,a,b} \psi_{q;k,ab} \ket{q;k,ab}.
\end{eqnarray}
Then, the particle density at $(R_1,i)$ is given as
\begin{align}
\langle \rho_{R_1,i} \rangle
 =& \bra{\Psi_q} \rho_{R_1,i} \ket{\Psi_q} \notag \\
 =&\sum_{kab} \sum_{k^\prime a^\prime b^\prime} \psi_{q;k,ab}  \psi^\ast_{q;k^\prime,a^\prime b^\prime} 
 \bra{q;k^\prime,a^\prime b^\prime} \rho_{R_1,i} \ket{q;k,ab} \notag \\
 =&\sum_{kab} \psi_{q;k,ab} \xi_{q;k,ab}(R_1,i),
\end{align}
where we have defined
\begin{align}
\xi_{q;kab} (R_1,i)= \sum_{k^\prime,a^\prime,b^\prime}\psi^\ast_{q;k^\prime,a^\prime b^\prime} 
 \bra{q;k^\prime,a^\prime b^\prime} \rho_{R_1,i} \ket{q;k,ab}.
\end{align}
The remaining task is to calculate $\xi_{q;kab}(R_1,i)$.
After some algebras with paying attention to the factor of $1/\sqrt{2}$ for the doubly-occupied states, we obtain
\begin{widetext}
\begin{align}
    &\xi_{q;k,ab}=\notag\\
    &\begin{cases}
    \sum_{a^\prime}[ \mathcal{N}_{k,a^\prime a} \psi^\ast_{q;k,a^\prime b} +
    \mathcal{N}_{q-k,a^\prime b} \psi^\ast_{q;k, a a^\prime}]
    &\mathrm{for~}
    k\not\equiv q-k~(\mathrm{Mod}~G)  \\
    \sum_{a^\prime \neq b} [\mathcal{N}_{k,a^\prime a} \psi^\ast_{q;k, ((a^\prime b))}] +\sum_{a^\prime \neq a}
    [ \mathcal{N}_{k,a^\prime b} \psi^\ast_{q;k, ((a a^\prime))}]\\
    +\sqrt{2} [
    \mathcal{N}_{k,ba} \psi^\ast_{q;k,bb}
    + \mathcal{N}_{k,ab} \psi^\ast_{q;k,aa}
    ] 
    & \mathrm{for~}
    k\equiv q-k~(\mathrm{Mod}~G)
    ~\mathrm{and}~a \neq b \\
    \sqrt{2} \sum_{a^\prime \neq a} [\mathcal{N}_{k,a^\prime a} \psi^\ast_{q;k ((a^\prime a))}]
    + 2\mathcal{N}_{k,aa} \psi^\ast_{q;k,aa}
    &\mathrm{for~}
    k\equiv q-k~(\mathrm{Mod}~G)
    ~\mathrm{and}~a=b
    \end{cases} \hspace{1mm}, \label{eq:twoparticle_xi}
\end{align}
\end{widetext}
where we have abbreviated $(R_1,i)$ for simplicity of writing, and the symbol $((ab))$ stands for the ascending ordering of $a$ and $b$. 

We note that in Fig.~\ref{fig3}(d) we plot the particle density per the unit cell at $R_1$ by taking the summation over the sublattice index $i$ for $\langle \rho_{R_1,i} \rangle$.
It is also worth noting that 
$\sum_{R_1,i}\langle \rho_{R_1,i} \rangle = 2$ holds, since we consider the two-particle systems.

\section{Explicit forms of tight-binding models \label{app:tbmodels}}
In this appendix, we provide the explicit forms of the 
tight-binding Hamiltonians studied in Sec.~\ref{sec:various_model}.
Note that we set the system size as $L_1 = L_2 = 16$ for all the models
except for the Haldane model;
for the Haldane model, we set $L_1 = L_2 = 24$ so that the Dirac points
of the two-particle band [Fig.~\ref{fig6}(a3)] become manifest. 

\subsection{2-band models}
\subsubsection{Qi-Wu-Zhang (QWZ) model}
This model was proposed in Ref.~\cite{Qi-Wu-Zhang-06} and is also referred to as the Wilson-Dirac model. 
The Bloch Hamiltonian reads
\begin{align}
    &H(\bm{k})=\sin k_1 \sigma_x+\sin k_2 \sigma_y+(m-\cos k_1-\cos k_2)\sigma_z, \label{eq:QWZ_Bloch}
\end{align}
where $\sigma_{x,y,z}$ are the Pauli matrices. 
The reciprocal lattice vectors are given as
\begin{align}
     \bm{G}_1=2\pi(1,0),\bm{G}_2=2\pi(0,1).
\end{align}
In numerical calculations, we set $m=1$.

\subsubsection{Modified QWZ model}
This model is given by simply swapping $\sigma_x$ for $\sigma_z$ in Eq.~(\ref{eq:QWZ_Bloch}). 
Specifically, the Bloch Hamiltonian reads
\begin{align}
    &H(\bm{k})=\sin k_1 \sigma_z+\sin k_2 \sigma_y+(m-\cos k_1-\cos k_2)\sigma_x.
\end{align}
In numerical calculations, we again set $m=1$.

\subsubsection{Haldane model}
The Haldane model was proposed as a lattice construction of the IQH system without an external magnetic field \cite{Haldane-88}.
We here write down the Haldane model with a slight modification that was introduced for the FCI study \cite{Neupert-Santos-Chamon-Mudry-11}.
Because this change only modifies the flatness of the lowest-band dispersion, it does not affect the calculations with projection. 
The Bloch Hamiltonian and the reciprocal lattice vectors are given as

\begin{align}
    &H(\bm{k})=b_01+\bm{b}\cdot\bm{\sigma},\\
    &\bm{G}_1=2\pi(1/\sqrt{3},1/3),
    \bm{G}_2=2\pi(0,2/3),
\end{align}
where 
\begin{align}
    &b_0=2t_2\cos \phi\left[\cos K_1+\cos K_2+\cos (K_1+K_2)\right],\notag\\
    &b_x=t_1\left[1+\cos(K_1+K_2)+\cos K_2\right],\notag\\
    &b_y=t_1\left[\sin (K_1+ K_2)  + \sin K_2\right],\notag\\
    &b_z=- 2t_2\sin\phi\left[\sin K_1+\sin K_2-\sin(K_1+K_2)\right]
\end{align}
with
\begin{align}
    &K_1=\sqrt{3} k_1, K_2=-\frac{\sqrt{3}}{2}k_1+\frac{3}{2}k_2.
\end{align}
We adopt the parameters, 
\begin{align}
    t_1=t_2=1/p_0, \phi=\cos^{-1} p_0,
\end{align}
where $p_0 =3\sqrt{3/43}$.

\subsubsection{Checkerboard model}
This model was studied in terms of the fermionic FCI \cite{Neupert-Santos-Chamon-Mudry-11}. 
The Bloch Hamiltonian and the reciprocal lattice vectors are given as
\begin{align}
    &H(\bm{k})=
    \begin{pmatrix}
    2t_2(\cos k_1-\cos k_2)&t_1f^*(\bm{k})\\
    t_1f(\bm{k})&-2t_2(\cos k_1-\cos k_2)\\
    \end{pmatrix},\\
    &\bm{G}_1=2\pi(1,0),\bm{G}_2=2\pi(0,1),
\end{align}
where
\begin{align}
    &f(\bm{k})=e^{-i\pi/4}\left[1+e^{i(k_2-k_1)}\right]+e^{i\pi/4}\left[e^{-ik_1}+e^{ik_2}\right],\notag\\
    &t_1=1,t_2=\sqrt{2}/2.
\end{align}

\subsection{$n_{\rm in}\geq$3-band models}

\subsubsection{Kagome model}
This model was studied in terms of the fermionic FCI \cite{Tang-Mei-Wen-11}.
There are three orbitals on different sublattices with position vectors:
\begin{align}
\bm{r}_1 = (0,0),\bm{r}_2 = \left(\frac{1}{2},0 \right),\bm{r}_3 = \left(\frac{1}{4},\frac{\sqrt{3}}{4} \right).
\end{align}
As we mentioned in the main text, there are two typical conventions for Fourier transform, depending on whether or not the relative positions $\bm{r}_i$ inside the unit cell are included in the Fourier factor.
The model in Ref. \onlinecite{Tang-Mei-Wen-11} was expressed in the convention that includes the information of positions:

\begin{widetext}
\begin{align}
&H_{\rm posi-dep}(\bm{k})=-2t_1
\begin{pmatrix}
0 & \cos K_1 & \cos K_2 \\
\cos K_1 & 0 & \cos K_3 \\
\cos K_2 & \cos K_3 & 0 \\
\end{pmatrix} + 2i \lambda_1 
\begin{pmatrix}
0 & \cos K_1 & -\cos K_2 \\
-\cos K_1 & 0 & \cos K_3 \\
\cos K_2 & -\cos K_3 & 0 \\
\end{pmatrix}\notag\\
&- 2t_2 \begin{pmatrix}
0 & \cos (K_2+K_3) & \cos (K_3-K_1) \\
\cos (K_2+K_3) & 0 & \cos (K_1+K_2) \\
\cos (K_3-K_1) & \cos (K_1+K_2) & 0 \\
\end{pmatrix} 
+ 2i\lambda_2
\begin{pmatrix}
0 & -\cos (K_2+K_3) & \cos (K_3-K_1) \\
\cos (K_2+K_3) & 0 & -\cos (K_1+K_2) \\
-\cos (K_3-K_1) & \cos (K_1+K_2) & 0 \\
\end{pmatrix}.
\end{align}
\end{widetext}
with
\begin{align}
K_1 = \bm{k} \cdot(\bm{r}_2-\bm{r}_1),
K_2 = \bm{k} \cdot(\bm{r}_3-\bm{r}_1),
K_3 = \bm{k}\cdot (\bm{r}_3-\bm{r}_2).
\end{align}
The two conventions are related via the following unitary transformation:
\begin{align}
H(\bm{k})  = D^\dagger_{\bm{k}} H_{\rm posi-dep}(\bm{k})D_{\bm{k}},
\end{align}
where
$ D_{\bm{k}} = \mathrm{diag} \left(e^{-i\bm{k} \cdot \bm{r}_1} ,e^{-i\bm{k} \cdot \bm{r}_2},e^{-i\bm{k} \cdot \bm{r}_3} \right)$.
The reciprocal lattice vectors are given as
\begin{align}
\bm{G}_1 = \left(2\pi, -2 \pi/\sqrt{3}\right),
\bm{G}_2 = \left( 0, 4\pi/\sqrt{3}\right).
\end{align}
We adopt the parameters 
$(t_1,t_2,\lambda_1,\lambda_2) = (1,-0.3, 0.28, 0.2)$.

\subsubsection{Square-lattice model}
This model was studied in terms of the fermionic FCI \cite{Sun-Gu-Katsura-DasSarma-11}.
This model is a three-orbital model on a square lattice. 
The Bloch Hamiltonian and reciprocal vectors are given as
\begin{widetext}
\begin{align}
    H(\bm{k}) = 
    \begin{pmatrix}
    -2t_{\rm dd} (\cos k_1 + \cos k_2) + \delta & 2i t_{\rm pd} \sin k_1 & 2it_{\rm pd} \sin k_2 \\
     & 2t_{\rm pp} \cos k_1 -2t_{\rm pp}^\prime \cos k_2 & i \Delta \\
      (\mathrm{h.c.})& & 2t_{\rm pp} \cos k_2 -2t_{\rm pp}^\prime \cos k_1 \\
    \end{pmatrix},\bm{G}_1 = (2\pi,0),\bm{G}_2 = (0,2\pi).
\end{align}
\end{widetext} 
We adopt the parameters, 
$t_{\rm dd} = t_{\rm pd} = t_{\rm pp} = 1$, 
$\Delta = 2.8$, 
$\delta = -4 t_{\rm dd} + 2 t_{\rm pp} + \Delta 
- 2t_{\rm pp} \Delta /  (4t_{\rm pp} + \Delta)$,
$t_{\rm pp}^\prime = t_{\rm pp} \Delta/ (4t_{\rm pp} + \Delta)$.

\subsubsection{Hofstadter ($1/4$-flux)}
The Hofstadter model was investigated as a lattice implementation of the Landau levels \cite{Hofstadter-76}.
For the number of sites per unit cell, $q$,
the Hofstadter model is a $q$-band model.
In this paper, we consider the four-band Hofstadter model.
The Bloch Hamiltonian and reciprocal vectors are given as
\begin{widetext}
\begin{align}
    H(\bm{k})=
    \begin{pmatrix}
    2\cos k_2&1&0&e^{-ik_1}\\
    1&2\cos (k_2+\pi/2)&1&0\\
    0&1&2\cos (k_2+\pi)&1\\
    e^{ik_1}&0&1&2\cos (k_2+3\pi/2)
    \end{pmatrix},~\bm{G}_1=2\pi(1,0),\bm{G}_2=2\pi(0,1).
\end{align}
\end{widetext}
\subsubsection{Ruby-lattice model}
This model was studied in terms of the fermionic FCI \cite{Hu-Kargarian-Fiete-11,Wu-Bernevig-Regnault-12}.
The Bloch Hamiltonian and reciprocal vectors are given as

\begin{widetext}
\begin{align}
&H(\bm{k})
= - \begin{pmatrix}
0  & & & & & \\
  \tilde{t}_1^\ast & 0 & & & & \\
  \tilde{t} & \tilde{t}_1^\ast e^{-i(K_1 + K_2)} & 0 & & & \\
  t_4 ( 1 + e^{i K_1}) & \tilde{t} & \tilde{t}_1^\ast e^{i K_1} & 0 & & \\
  \tilde{t}^\ast & t_4( 1 + e^{-i (K_1+K_2)}) & \tilde{t} & \tilde{t}_1^\ast & 0 & \\
  \tilde{t}_1e^{iK_1} & \tilde{t}^\ast & t_4 ( e^{iK_1} + e^{i (K_1+K_2)}) & \tilde{t} & \tilde{t}_1^\ast e^{i (K_1+K_2)} & 0 \\
  \end{pmatrix}
  + (\mathrm{h.c.}), \notag\\
  &\bm{G}_1 = (2\pi, 2\pi/\sqrt{3}),\bm{G}_2 = (0,4\pi /\sqrt{3}),
\end{align}
\end{widetext}
where
$\tilde{t} = t_r + it_i$, $\tilde{t}_1 = t_{1r} + i t_{1i}$, $K_1 = k_1$, $K_2 = -k_1/2 + \sqrt{3}k_2/2$.
We adopt the parameters $(t_r,t_i,t_{1r},t_{1i},t_4) = (1,1.2, -1.2, 2.6, -1.2)$.

\bibliography{FCI}
\end{document}